\documentclass[structabstract]{aa}
\usepackage{txfonts}
\usepackage{natbib}
\usepackage{graphicx}
\usepackage{aalongtable}
\bibpunct{(}{)}{;}{a}{}{,}

\begin{document}

\title{CoRoT high-precision photometry of the B0.5\,IV star HD\,51756 \thanks{The CoRoT space mission was developed and is operated by the French space agency CNES, with participation of ESA's RSSD and Science Programmes, Austria, Belgium, Brazil, Germany, and Spain.}\fnmsep\thanks{Based on data gathered with \textsc{Coralie} installed on the 1.2 meter Euler telescope at La Silla, Chile; and \textsc{Harps} installed on the 3.6 meter ESO telescope (ESO Large Programme 182.D-0356) at La Silla, Chile.}}

\author{P.~I.~P\'{a}pics\inst{\ref{inst1}} 
\and M.~Briquet\inst{\ref{inst1}}\thanks{Postdoctoral Fellow of the Fund for Scientific Research, Flanders.} 
\and M.~Auvergne\inst{\ref{inst2}}
\and C.~Aerts\inst{\ref{inst1},\ref{inst3}} 
\and P.~Degroote\inst{\ref{inst1}} 
\and E.~Niemczura\inst{\ref{inst1},\ref{inst4}} 
\and M.~Vu\v{c}kovi\'{c}\inst{\ref{inst1},\ref{inst5}} 
\and K.~Smolders\inst{\ref{inst1}}\thanks{Aspirant Fellow of the Fund for Scientific Research, Flanders.}
\and E.~Poretti\inst{\ref{inst8}}
\and M.~Rainer\inst{\ref{inst8}}
\and M.~Hareter\inst{\ref{inst9}}
\and A.~Baglin\inst{\ref{inst10}} 
\and F.~Baudin\inst{\ref{inst6}} 
\and C.~Catala\inst{\ref{inst2}}
\and E.~Michel\inst{\ref{inst2}}
\and R.~Samadi\inst{\ref{inst2}}}

\institute{Instituut voor Sterrenkunde, K.U.Leuven, Celestijnenlaan 200D, B-3001 Leuven, Belgium\label{inst1} 
\and LESIA, UMR8109, Universit\'{e} Pierre et Marie Curie, Universit\'{e} Denis Diderot, Observatoire de Paris, 92195 Meudon Cedex, France\label{inst2}
\and Department of Astrophysics, IMAPP, University of Nijmegen, PO Box 9010, 6500 GL Nijmegen, The Netherlands\label{inst3}
\and Instytut Astronomiczny, Uniwersytet Wroclawski, Kopernika 11, 51-622, Wroclaw, Poland\label{inst4} 
\and European Southern Observatory (ESO), Alonso de C\'{o}rdova 3107, Vitacura, Cassilla 19001, Santiago, Chile\label{inst5} 
\and INAF - Osservatorio Astronomico di Brera, via E.~Bianchi 46, 23807 Merate (LC), Italy\label{inst8}
\and Institut f\"{u}r Astronomie, Universit\"{a}t Wien, T\"{u}rkenschanzstrasse 17, A-1180 Vienna, Austria\label{inst9}
\and Laboratoire AIM, CEA/DSM-CNRS-Universit\'{e} Paris Diderot; CEA, IRFU, SAp, centre de Saclay, F-91191, Gif-sur-Yvette, France\label{inst10} 
\and Institut d'Astrophysique Spatiale, CNRS/Univ. Paris-Sud, B\^{a}t. 121, F-91405, Orsay Cedex, France\label{inst6}} 

\date{Received ? ???? 2010 / Accepted ? ???? 2010}

\abstract {OB stars are important constituents for the ecology of the Universe, and there are only a few studies on their pulsational properties detailed enough to provide important feedback on current evolutionary models.}{Our goal is to analyse and interpret the behaviour present in the CoRoT light curve of the B0.5\,IV star HD\,51756 observed during the second long run of the space mission, and to determine the fundamental stellar parameters from ground-based spectroscopy gathered with the \textsc{Coralie} and \textsc{Harps} instruments after checking for signs of variability and binarity, thus making a step further in mapping the top of the $\beta$\,Cep instability strip.}{We compare the newly obtained high-resolution spectra with synthetic spectra of late O-type and early B-type stars computed on a grid of stellar parameters. We match the results with evolutionary tracks to estimate stellar parameters. We use various time series analysis tools to explore the nature of the variations present in the light curve. Additional calculations are carried out based on distance and historical position measurements of the components to impose constraints on the binary orbit.}{We find that HD\,51756 is a wide binary with both a slow ($v \sin i \approx 28\,\mathrm{km\,s}^{-1}$) and a fast ($v \sin i \approx 170\,\mathrm{km\,s}^{-1}$) early-B rotator whose atmospheric parameters are similar ($T_\mathrm{eff}\approx30000\,\mathrm{K}$ and $\log g \approx 3.75$). We are unable to detect pulsation in any of the components, and we interpret the harmonic structure in the frequency spectrum as sign of rotational modulation, which is compatible with the observed and deduced stellar parameters of both components.}{The non-detection of pulsation modes provides a feedback on the theoretical treatment, given that non-adiabatic computations applied to appropriate stellar models predict the excitation of both pressure and gravity modes for the fundamental parameters of this star.}


\maketitle


\section{Introduction}\label{intro}
One of the major observing programmes for the asteroseismology channel of the CoRoT satellite \citep{2006ESASP1306...33B}, is to derive details of the internal physics of stars within the instability strip of early-type B pulsators \citep[e.g.,][]{2006ESASP1306...39M}.  A major motivation of this programme was to understand the excitation problems reported for some of the observed modes in several prototypical bright $\beta$\,Cep stars by groups using independent stellar evolution and pulsation codes \citep[e.g.,][]{2004MNRAS.350.1022P, 2004MNRAS.355..352A, 2006MNRAS.365..327H, 2009MNRAS.396.1460D}, while other class members show oscillations as predicted by theory \citep{2003Sci...300.1926A, 2006A&A...459..589M, 2007MNRAS.381.1482B, 2009MNRAS.398.1339H}. The goal was also to study stars of spectral type O or B0, which are higher up in the predicted $\beta$\,Cep strip than most class members observed from the ground prior to CoRoT, to see if the emptiness of that part of the strip is removed when going from mmag to $\mu$mag precision in photometric data.  A third motivation was to obtain data of higher precision than from the ground and uninterrupted, in order to establish once and for all the firm detection of non-radial gravity modes seemingly present in ground-based spectroscopy of rapidly rotating Be stars \citep[e.g.,][]{2003A&A...411..229R}, and to interpret those in terms of oscillations of rapid rotators. A number of early-type B stars were thus selected to fulfill these goals, keeping in mind the strong pointing restrictions of the satellite as well as the brightness limitations \citep[e.g.,][]{2009A&A...506..411A}. The star HD\,51756 (B0.5\,IV) studied in this paper was selected as target in this context, as well as 6 O stars measured during a short run. All further planned B targets have later spectral type due to pointing restrictions of the satellite. 

The slowly rotating B stars observed by CoRoT's asteroseismology CCDs so far, have led to some remarkable results. The only accessible known $\beta$\,Cep star, HD\,180642 (B1.5\,II-III), turned out to have numerous self-excited modes rather than one dominant radial mode as was thought prior to CoRoT \citep{2009A&A...506..111D,2009A&A...506..269B} and many of the observed modes are resonantly coupled. Moreover, the star was found to have modes of a stochastic nature as well \citep{2009Sci...324.1540B}, which was never found before in a $\beta$\,Cep pulsator and which requires the presence of a surface convection zone capable to excite such modes. Stochastically excited modes exhibiting a constant frequency spacing were, surprisingly, detected in the O8.5\,V star HD\,46149 \citep{2010A&A...519A..38D}. On the other hand, the O9V star HD\,46202 turns out to be the most massive star known to date with $\beta$\,Cep-like pulsation frequencies, but for which present excitation computations fail to predict the observed oscillations \citep{Briquet2010}. Further, at the cool border of the $\beta$\,Cep instability strip, the B3\,V star HD\,50230 revealed oscillations of hybrid nature, i.e., the presence of both gravity modes as expected in slowly pulsating B stars and pressure modes as in the classical $\beta$\,Cep stars. The gravity modes of this star displayed a period spacing and a periodic deviation thereof, which allowed one to prove the existence of a chemically inhomogeneous zone around the fully mixed core which is not yet taken into account in current B star models \citep{2010Natur.464..259D}.

As for the rapidly rotating early-Be stars, HD\,49330 (B0.5\,IVe) turned out to be a clear case where multiple mode beating was observed in real time and this phenomenon was found to be connected with an outburst feeding the circumstellar disk \citep{2009A&A...506...95H, 2009A&A...506..103F}. This supports the early findings of \citet{1988A&A...198..211B} and \citet{2001A&A...369.1058R} that the outbursts detected in the B2e star $\mu$\,Cen have a pulsational origin and are connected with multimode beating. This outburst phenomenon was so far not found for the other Be stars which were monitored intensively by CoRoT, but those targets are of later spectral type (\citealt[HD\,181231, B5\,IVe]{2009A&A...506..143N}; \citealt[HD\,175869, B8\,IIIe]{2009A&A...506..133G}). There is yet no acceptable seismic modeling of the interior structure of a Be star, due to the mathematical complexity of the treatment of stellar oscillations in a deformed star.

It is obvious from these recent findings that the diversity of stellar oscillations in and near the $\beta$\,Cep instability strip is much
larger than anticipated before the CoRoT mission. This must imply that details in the internal physics of the various stars, such as their internal rotation, mixing, settling and radiative levitation due to atomic diffusion, etc., may be different. Some problems may also be solved by increasing the opacity in the excitation layers. In an attempt to increase the number of well studied early-B stars, HD\,51756 (B0.5\,IV) was monitored by CoRoT's seismology CCDs and studied in parallel with ground-based spectroscopy.


\section{Fundamental parameters}\label{fundparameters}

\subsection{Prior to our study}\label{prior}
HD\,51756 -- a bright ($V_\mathrm{mag}=7.23$) B0.5\,IV field-star in the constellation of Monoceros -- was first listed among early-type high-luminosity objects by \citet{1951ApJ...113..309M}, being classified as a B3 star from objective-prism plates. The determination of the luminosity class and a more precise spectral type of B0.5\,IV was done by \citet{1955ApJS....2...41M}, which is generally used since then. Although there are exceptions: a B0 III classification by \citet{1982MNRAS.200..445D}, a B1\,V by \citet{2008ApJS..176..216A}, and a B1/2\,Ib by \citet{2003AJ....125..359W} -- there seems to be a quite good agreement on the spectral type, but not the luminosity class.

Measurements by \citet{1875AN.....86..337B} showed that HD 51756 is a triple system (catalogued as BU\,327, but widely known as WDS\,J06585-0301), consisting of two almost equally bright ($\Delta m\approx0.3$ mag) close components with a separation of $\rho\approx0.7$\arcsec, and a fainter star further away ($\Delta m\approx4$ mag, $\rho\approx13.7$\arcsec). Unfortunately, only one historical radial velocity (RV) measurement ($v_\mathrm{rad}=25\pm5\,\mathrm{km\,s}^{-1}$) can be found in the literature \citep{1953GCRV..C......0W}. Searching for $\beta$\,Cep stars, \citet{1967ApJS...14..263H} took the first photometric time-series and concluded that this field-star was constant. Among other parameters, he also measured the projected rotational velocity to be $v \sin i = 30\,\mathrm{km\,s}^{-1}$. There is also a measurement from \citet{2000AcA....50..509G} claiming a $v \sin i = 40\,\mathrm{km\,s}^{-1}$.

Further information we found concerns $V_\mathrm{mag}=7.23$, $\beta_\mathrm{mag}=2.569$, $M_\mathrm{v}=-4.1$, the colours of $(B-V)=-0.10$ and $(U-B)=-0.93$, and the colour excess of $E(B-V)=0.19$. In addition to these, $(UV\footnote{the $UV$ filter is centered at $1500\AA$}-V)=-2.60$ and $E(UV)=1.07$ was measured by \citet{1971ApJ...166..543W}. A distance modulus of $V_\mathrm{0}-M_\mathrm{v}=10.7$ is given by \citet{1976A&A....53....9V}, which leads to a distance of 1.38\,kpc (1.04\,kpc with extinction taken into account). There are similar measurements of $1.3\pm0.5$\,kpc \citep[for the BFS\,55 \ion{H}{ii} region\footnote{for the same region \citet{1984ApJ...279..125F} found a kinematic distance of $1.89\pm0.73$\,kpc determined by using CO velocities of the associated molecular clouds and the CO rotation curve of the outer galaxy}, which is claimed to be ionised by the nearby multiple subgiant, HD 51756]{1984NInfo..56...59A}, 1.6\,kpc \citep{1985ApJS...59..397S} and $2.156^{+0.559}_{-0.443}$\,kpc \citep[with $M_\mathrm{v}=-5.075$ from $uvby\beta$ photometry]{2000MNRAS.312..753K}. The original Hipparcos \citep{1997A&A...323L..49P} parallax is $0.56\pm4.8$\,mas, while the value from the new reduction \citep{2007A&A...474..653V} is $-1.94\pm1.13$\,mas. Unfortunately none of these lead to a useful distance estimate.

Empirical temperature calibrations \citep{1989A&AS...80...73G} gave values of $\log T_\mathrm{eff}=4.522\pm0.051\,\mathrm{K}$, $\log T_\mathrm{eff}=4.532\pm0.042\,\mathrm{K}$, and $\log T_\mathrm{eff}=4.530\pm0.045\,\mathrm{K}$, derived from QUV, $(m1965-V)_0$, and $(B-V)_0$, respectively (corresponding to -- with errors taken into account -- an effective temperature range of $29\,500-37\,500$\,K).


\subsection{New spectroscopy}

In the framework of ground based preparatory and follow-up observations for the CoRoT space mission, high resolution and medium to high signal to noise spectra were taken with the \textsc{Coralie} \citep[$R\approx50\,000$]{1996A&AS..119..373B,2001Msngr.105....1Q} and \textsc{Harps} \citep[$R\approx80\,000$]{2003Msngr.114...20M} instruments (78 and 15 measurements respectively -- see Table\,\ref{specsummary} for a summary of the spectroscopic observations). In addition to these, one spectrum gathered with \textsc{Feros} \citep[$R\approx48\,000$]{1999Msngr..95....8K} was taken from the GAUDI archive \citep{2005AJ....129..547S}.

\begin{table*}
\caption{Logbook of the new spectroscopic observations of HD\,51756 obtained between January 2009 and December 2009 -- grouped by observing runs.}
\label{specsummary}
\centering
\begin{tabular}{c c c c c c c c}
\hline\hline
Instrument & N & HJD begin & HJD end & $\langle\mathrm{SNR}\rangle$ & SNR-range & $\mathrm{T}_{\mathrm{exp}}$ & R\\
\hline
\textsc{Coralie} & 53 & 2454842 & 2454852 & 124 & [97, 142] & 1500        &$50\,000$\\
\textsc{Coralie} & 25 & 2454912 & 2454922 & 130 & [99, 160] & [1500, 2000]&$50\,000$\\
\textsc{Harps}   & 15 & 2455174 & 2455195 & 293 & [265, 330]& [450, 840]  &$80\,000$\\
\hline
\end{tabular}
\tablefoot{For each observing run, the instrument, the number of spectra N, the HJD, the average signal-to-noise ratio (calculated in the line free region of [5370\AA, 5400\AA]), the range of SNR values, the typical exposure times (in seconds), and the resolution of the spectrograph are given.}
\end{table*}

We performed a careful normalisation of each spectrum, using cubic splines which were fitted through some tens of points at fixed wavelengths, where the continuum was known to be free of spectral lines and telluric features. The final selection of these nodal points was done after several tests and quality checks, making sure that, e.g., the wings of the Balmer-lines are not affected, but all artificial features (like periodic waves in the continuum) are corrected by the process. Cosmic removal via pixel-by-pixel sigma clipping and a sophisticated order-merging (taking into account the signal to noise values in the overlapping ranges, and correcting for the sometimes slightly different flux-levels of the overlapping orders) was also done by the same semi-automatic script. This is much faster and less subjective than one-by-one manual normalisation.

\begin{figure}
\resizebox{\hsize}{!}{\includegraphics{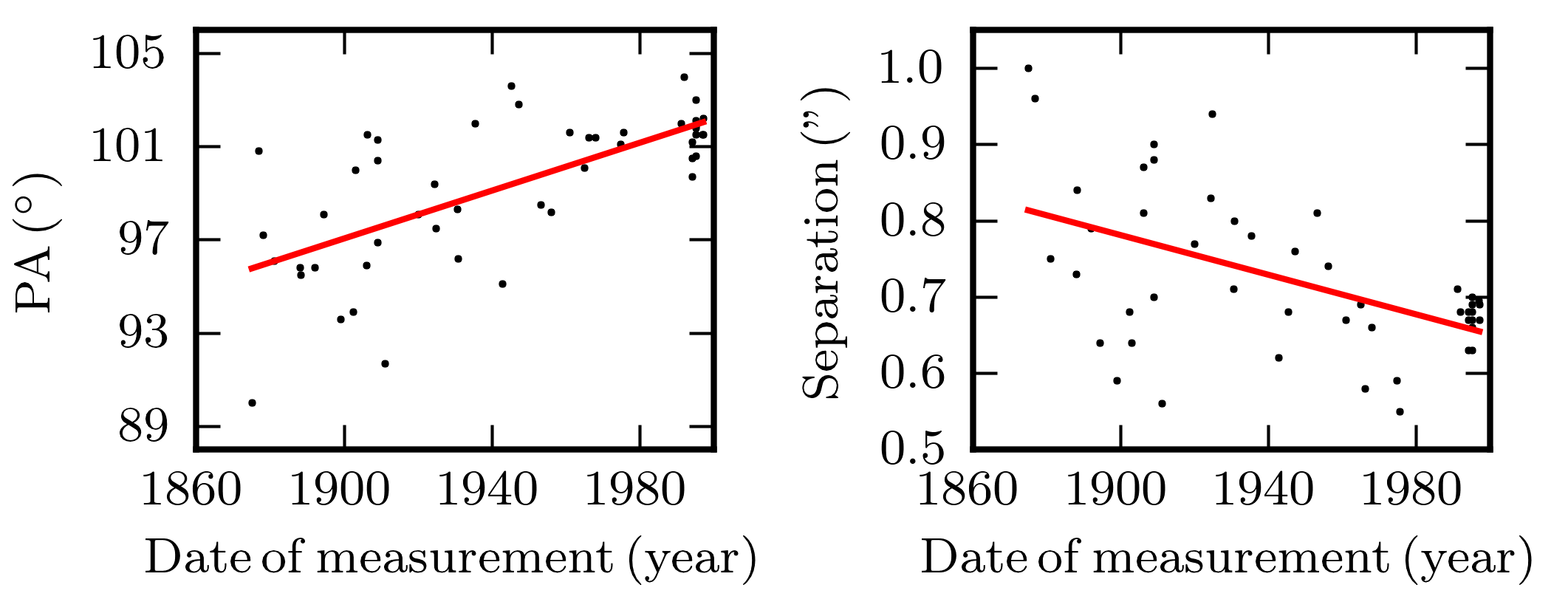}} 
\caption{Position angle (\textit{left}) and separation (\textit{right}) measurements of the two bright and close components of HD\,51756 (WDS\,J06585-0301\,AB) from The Washington Double Star Catalog, with a linear trend fitted to them.}
\label{wd_pa_sep}
\end{figure}

Radial velocities were measured from cross-correlated profiles -- which were calculated from the three strongest \ion{Si}{iii} lines (at $4553\AA$, $4568\AA$, and $4575\AA$) using the least-square deconvolution technique of \citet{1997MNRAS.291..658D} -- by fitting a simple Gaussian to them. It is clear from the determined radial velocities, that the orbital period has a timescale much longer than the timespan of the spectroscopic observations, and the orbit can not be fitted on this number of measurements. The idea of a very long orbital period is confirmed by position measurements covering 122 years found in the Washington Double Star Catalog also (see Table\,\ref{positions}), which show only minor change during more than a century in position angle (which slowly increases), and in separation (which slowly decreases) -- see Fig.\,\ref{wd_pa_sep}.

By visual inspection of the averaged spectrum of both instruments (and the original spectra), we conclude that HD\,51756 is a double-line spectroscopic binary (SB2) with a slow (referred to as the primary from now on, as the sharp spectral features are connected to this star) and a fast rotator (secondary) star (see Fig.\,\ref{syntheticfits}). Seeing both components in the spectrum is not surprising as the separation of the components is only $0.7\arcsec$, which is smaller than the fiber aperture of \textsc{Coralie} ($2\arcsec$) and \textsc{Harps} ($1\arcsec$), and it is also below the average seeing of La Silla ($0.9\arcsec$), while the difference in brightness of the primary and the secondary is small.

Knowing the RV values corresponding to each exposure, we transferred all spectra to the laboratory rest frame (of the primary). At this point, we re-run the normalisation script (this way the nodal points are really at the same rest wavelength), and calculated the average spectrum for both instruments -- using weights according to the calculated or given SNR values of each spectrum. The two average spectra calculated from \textsc{Coralie} and \textsc{Harps} are almost perfectly identical (with an SNR [5370-5400\AA] of 873 and 1016, respectively). The spectra show only absorption features, except the \ion{Fe}{iii} emission line at $5243\AA$.

Though in the case of the individual \textsc{Coralie} spectra, the SNR levels were a bit lower than one would typically have for such analysis, we searched for line-profile variations using the FAMIAS software package \citep{2008CoAst.155...17Z}, but without any positive detection.


\subsection{Fundamental parameters}\label{gridsearch}

An SB2 binary is extremely useful when the orbit can be fitted and, after spectral disentangling, a full analysis can be carried out on both components separately. In our case, this is not possible, due to the lack of knowledge concerning the orbital parameters.

\begin{figure*}
\resizebox{\hsize}{!}{\includegraphics{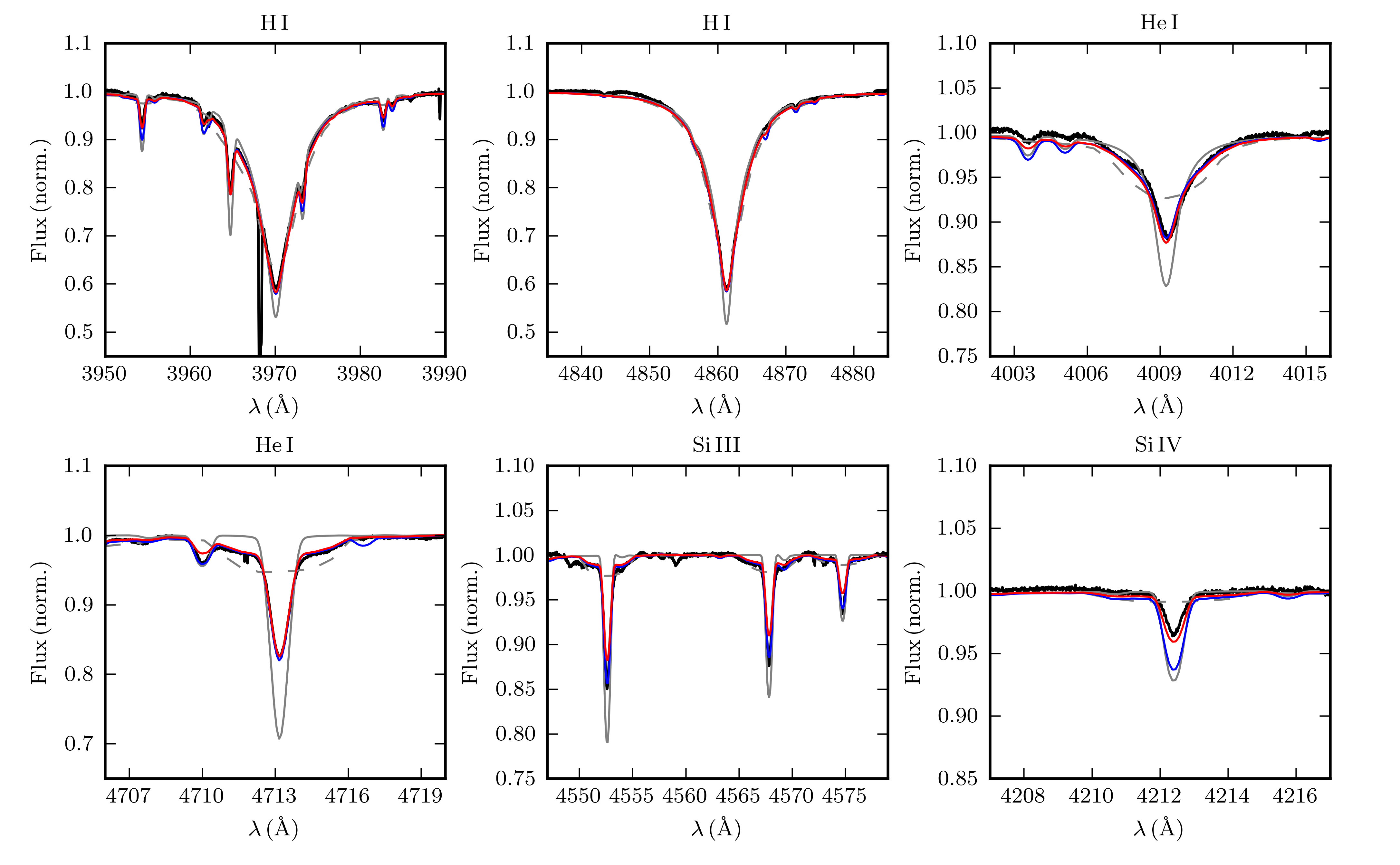}} 
\caption{Fits to the averaged \textsc{Harps} spectrum (black). The parameters are listed in Table\,\ref{gridsearchresults}. The fit to the primary companion is shown in solid grey, the fit to the secondary in dashed grey, and the combined fit in red. While these synthetic spectra were calculated for $Z=0.010$, we also plot the combined spectrum corresponding to $Z=0.020$ with a solid blue line to show the effect of the change in metallicity. The effect of the fast rotator companion is clearly visible in the \ion{He}{i} line at $4713\AA$ (\textit{lower left panel}). We had difficulties in fitting the metal lines equally well with any given setting. Some regions suggest higher (e.g., the \ion{Si}{iii} lines at $4553\AA$, $4568\AA$, and $4575\AA$ (\textit{lower middle})), while some others suggest lower metallicity values (e.g., the \ion{Si}{iv} line at $4212\AA$ (\textit{lower right})).}
\label{syntheticfits}
\end{figure*}

To have a relatively fast but still accurate solution, we decided to carry out a full grid search with seven free parameters ($T_\mathrm{eff}$, $\log g$, and  $v \sin i$ for both components, plus the metallicity $Z$ which was assumed to be the same for both stars), using the OSTAR2002 \citep{2003ApJS..146..417L} and BSTAR2006 \citep{2007ApJS..169...83L} grids. These sets of synthetic spectra were both calculated from NLTE, plane-parallel, hydrostatic model atmospheres, with a microturbulent velocity of $10\,\mathrm{km\,s}^{-1}$ and $2\,\mathrm{km\,s}^{-1}$ for the OSTAR2002 and BSTAR2006 grid, respectively. Equivalent widths of the instrumental line profiles were also taken into account while calculating the broadening of the synthetic composite spectra. Chi-square ($\chi^{2}$) values were calculated using only regions of selected \ion{H}{i}, \ion{He}{i}, \ion{Si}{iii}, and \ion{Si}{iv} lines (some example fits are shown in Fig.\,\ref{syntheticfits}). Errors were estimated by running the grid search using both the \textsc{Coralie} and \textsc{Harps} average spectra, using all four possible combinations of the two grids for the two components, and using more or fewer line regions for the $\chi^{2}$ calculation, then finally taking the most common parameters among the combinations with the lowest $\chi^{2}$ values as the final result, while the lowest and highest parameter values close to the minima of the $\chi^{2}$ set the error bars. A summary of the results is shown in Table\,\ref{gridsearchresults}. The two stars are very similar, the only difference being a different equatorial velocity or a different stellar inclination. The errors on the effective temperature and on the surface gravity might be slightly underestimated in case of the secondary, as this component contributes less to the overall appearance of the composite spectra because of the shallow lines caused by its high projected rotational velocity. The lowest $\chi^{2}$ values were reached by using the OSTAR2002 grid for the primary and the BSTAR2006 grid for the secondary (note that, not only the microturbulent velocity, but also the atomic data included in these model atmosphere grids are slightly different). Also a slight trend in the full set of results suggests that the secondary might be $\approx1000$\,K cooler than the primary and it might have a slightly lower $\log g$ value compared to its companion star, but these deviations are within the given error bars. As shown below, this is in agreement with the observed $\Delta m = m_2 - m_1 = M_2 - M_1 \approx0.3$ mag brightness difference of the components. From the definition of the magnitude scale, and the relation between effective temperature and luminosity, we write \[ \left( \frac{R_1}{R_2}\right)^2 \left( \frac{T_1}{T_2}\right)^4 = \frac{L_1}{L_2} = 10^{(M_2-M_1)/2.5} \approx 1.3\] and, using this expression, we can give the upper limit of the difference in effective temperature and radii of the components. Assuming the same radii, we need $T_2 = 0.937\,T_1$ to fulfill the equation, while assuming the same temperature requires $R_2 = 0.877\,R_1$. At a $T_{\mathrm{eff}}=30\,000\,\mathrm{K}$ this means that the secondary can be cooler by a maximum of $\approx2000\,\mathrm{K}$, which fits the results of the grid search within the errors.

\begin{table}
\caption{Observed and corresponding model parameters}
\label{gridsearchresults}
\centering
\renewcommand{\arraystretch}{1.25}
\begin{tabular}{c c c}
\hline\hline
Parameter & Primary & Secondary \\
\hline
\multicolumn{3}{c}{from spectroscopy} \\
\hline
$T_\mathrm{eff}\,(\mathrm{K})$ & $30000\pm1000$ & $30000\pm2500$\\
$\log g\,\mathrm{(cgs)}$ & $3.75\pm0.25$ & $3.75\pm0.25$ \\
$v \sin i\,(\mathrm{km\,s}^{-1})$ & $28\pm4$ & $170\pm15$ \\
$Z$ &\multicolumn{2}{c}{$\in[0.010, 0.020]$}\\
\hline
\multicolumn{3}{c}{from evolutionary tracks}\\
\hline
$\mathcal{M}\,(\mathcal{M}_{\sun})$&$18.9^{+9.1}_{-4.4}$&$18.9^{+15.1}_{-5.9}$\\
$R\,(R_{\sun})$&$9.6^{+5.9}_{-3.3}$&$9.6^{+7.5}_{-3.7}$\\
$\log L\,(L_{\sun})$&$4.8^{+0.5}_{-0.4}$&$4.8^{+0.6}_{-0.6}$\\
$\mathrm{age\,(Myr)}$&$7.6^{+1.1}_{-2.0}$&$7.6^{+2.8}_{-3.0}$\\
\hline
\end{tabular}
\end{table}

\begin{figure}
\resizebox{\hsize}{!}{\includegraphics{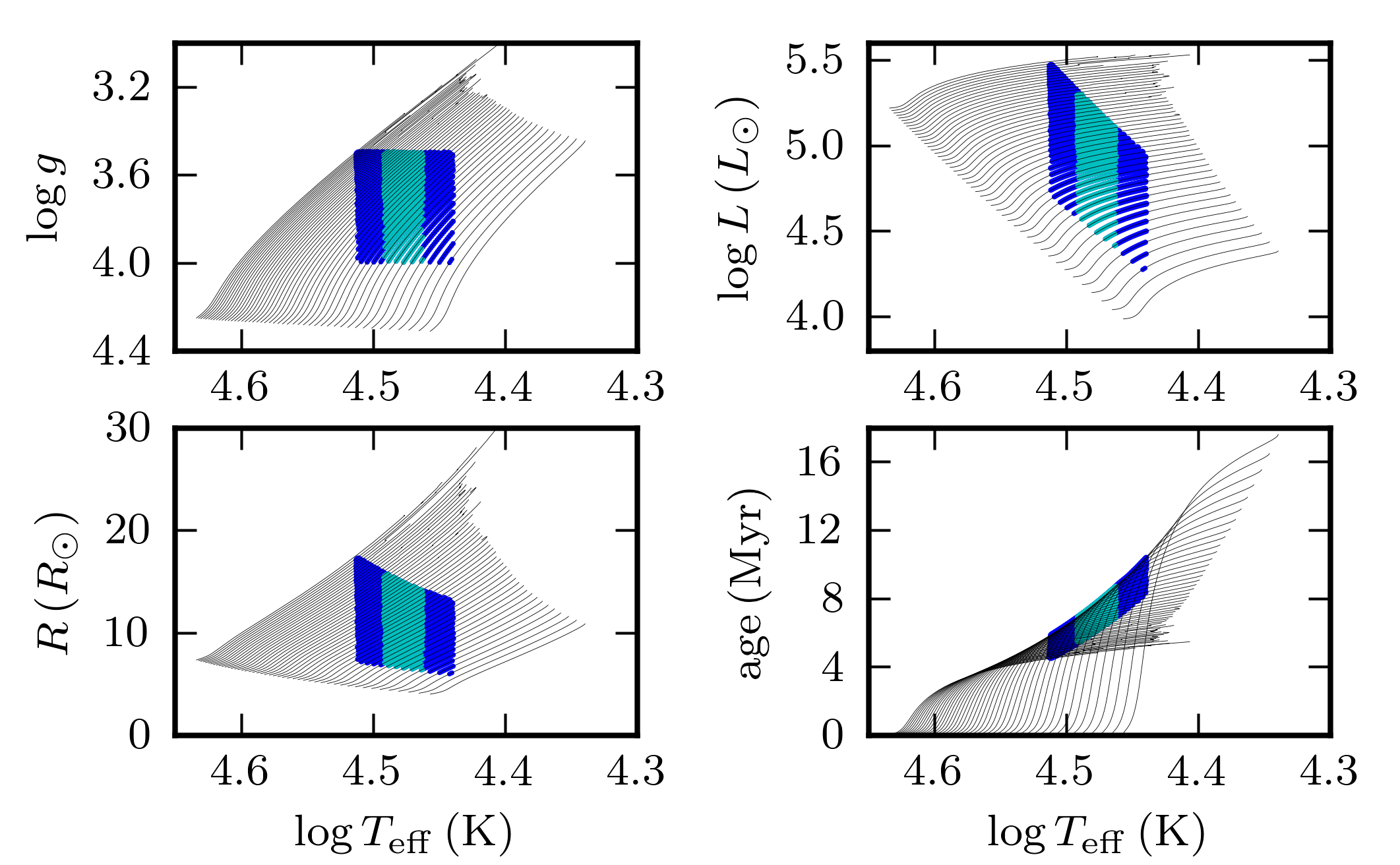}} 
\caption{Overview of the main sequence evolutionary tracks of late O-type and early B-type stars for masses of $12-35\mathcal{M}_{\sun}$ (with a step of $0.5\mathcal{M}_{\sun}$) and a core overshoot parameter $\alpha_\mathrm{ov}=0.2$  pressure scale heights. Grid-points which fall within the error boxes of the primary and secondary components are shown with cyan and blue shading, respectively, on the $\log T_\mathrm{eff}$--$\log g$ (\textit{upper left}), $\log T_\mathrm{eff}$--$\log L$ (\textit{upper right}), $\log T_\mathrm{eff}$--$R$ (\textit{lower left}), and $\log T_\mathrm{eff}$--age (\textit{lower right}) diagrams.}
\label{grids1}
\end{figure}

\begin{figure}
\resizebox{\hsize}{!}{\includegraphics{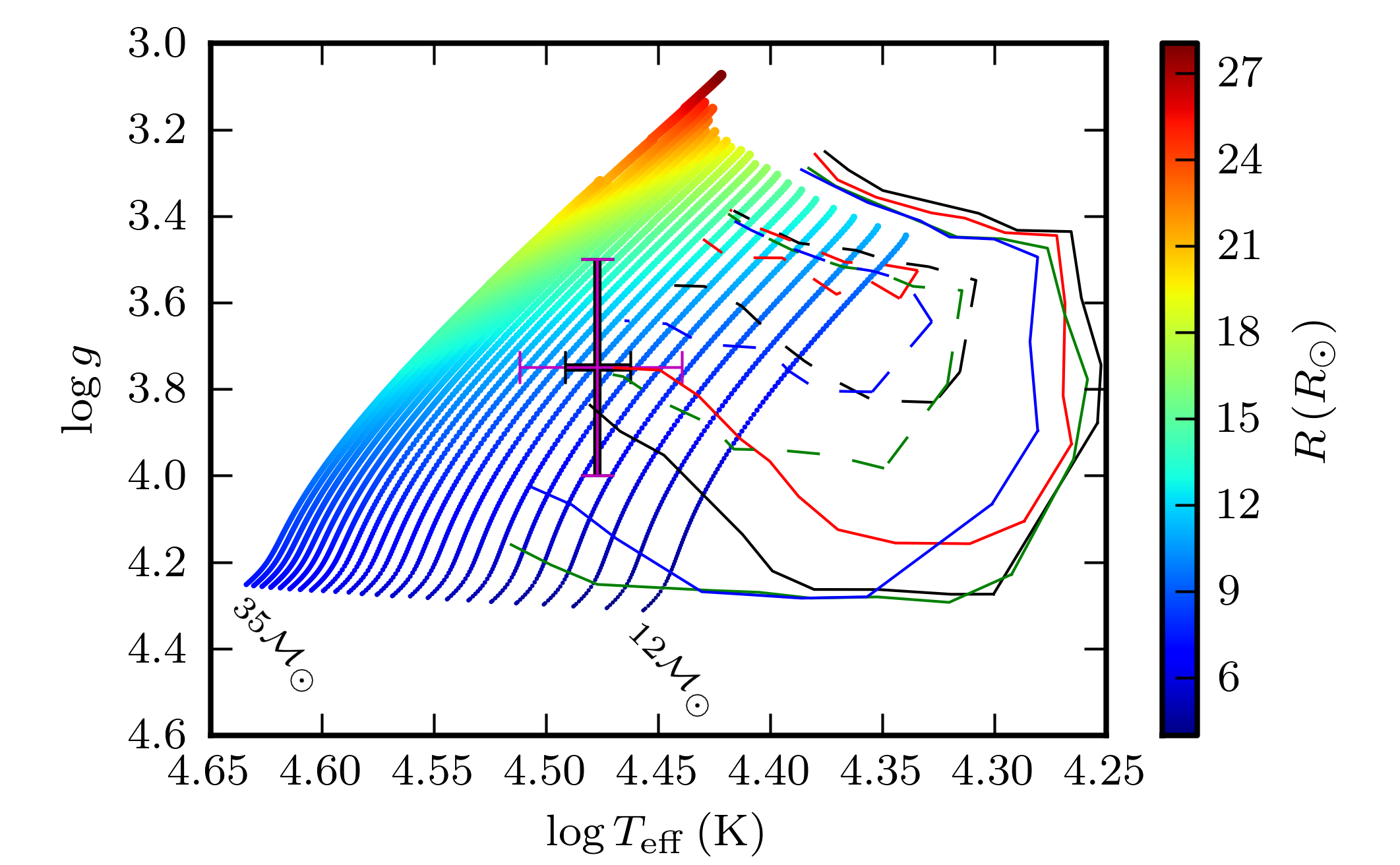}} 
\caption{$\log T_\mathrm{eff}$--$\log g$ diagram of late O-type and early B-type stars. The spectroscopic parameters and corresponding error estimations are indicated by thick and thin crosses for the primary and secondary component, respectively. The main sequence evolutionary tracks are for masses of $12-35\mathcal{M}_{\sun}$ (with a step of $1\mathcal{M}_{\sun}$) and a core overshoot parameter $\alpha_\mathrm{ov}=0.2$  pressure scale heights, and their shading and thickness indicates the corresponding radius values. Various $\beta$\,Cep instability strips for $Z=0.020$ (continuous lines) and $Z=0.010$ (dashed lines), both for four different combinations of metal mixtures and opacity computations (OPAL GN93, OP GN93, OPAL AGS05+Ne, and OP AGS05+Ne) are plotted with red, blue, black, and green colours, respectively. These are taken from \citet{2007CoAst.151...48M} who computed them for masses up to $18\mathcal{M}_{\sun}$.}
\label{grids2}
\end{figure}

After we determined the $T_\mathrm{eff}$ and $\log g$ values of both components, we matched them with evolutionary tracks (see Fig.\,\ref{grids1} and Fig.\,\ref{grids2} for an overview, and \citet{Briquet2010} for a description of the input physics). As the lowest $\chi^{2}$ values indicated a minimum in between $Z=0.02$ and $Z=0.01$, we fixed the metallicity at $Z=0.014$ in accordance with \citet{2008ApJ...688L.103P} who proposed the cosmic abundance standard (CAS) from early B-type stars in the solar neighborhood. We choose $X=0.715$ also in good agreement with the proposed CAS, and set the core overshoot parameter at $\alpha_{\mathrm{ov}}=0.2$ pressure scale heights, since asteroseismic modeling results of $\beta$\,Cep targets have given evidence for the occurrence of core overshooting of that order \citep{2010aste.book.....A}. From the model points which lie within the error box set by the $T_\mathrm{eff}$ and $\log g$ determination, we estimated the stellar parameters of the components. These values and their uncertainties are listed in Table\,\ref{gridsearchresults}.

\subsection{Constraints on the orbit}\label{orbit}

Using the masses from the evolutionary tracks (see Table\,\ref{gridsearchresults}), the distance-values (see Sect.\,\ref{prior}), and the visual separation from the position measurements (Table\,\ref{positions}), we calculate the expected orbital period of the AB components of the system to see if the slow change in position angle and separation is compatible with it. Assuming the simplest case, a circular, pole-on orbit, the semi-major axis (hence the radius in this simplified case) of the orbit is $a=0.0056^{+0.0036}_{-0.0030}$\,pc (using a distance of $d = 1.7^{+1.0}_{-0.9}$\,kpc representing the values mentioned in Sect.\,\ref{prior}, and a separation of $\rho=0.68\pm0.2\arcsec$ calculated from the most recent and precise interferometric measurements between 1994 and 1997). Using 
\[P = 2\pi\sqrt{\frac{a^3}{G\left(\mathcal{M}_\mathrm{A}+\mathcal{M}_\mathrm{B}\right)}}\]
we obtain an orbital period of $P = 6385^{+9379}_{-4808}$\,year -- with the errors covering all uncertainties. These values suggest a change of $6.9^{+21.0}_{-4.1}$\,degrees in position angle during  122 years, which fits the observations (showing a change of $\approx6.3\degr$). Assuming an edge-on orbit, we derive an upper limit of the semi-amplitude of the radial velocity; $A_\mathrm{RV}= 5.4^{+4.7}_{-1.8}\,\mathrm{km\,s}^{-1}$. This -- combined with the shallow profiles of the secondary due to its high projected rotational velocity and the relatively short timespan of the measurements (compared to the orbital period) -- explains why there was no (measurable) wavelength-shift between the lines of the two components in the spectra.






\section{The CoRoT light curve}

\subsection{Frequency analysis}\label{frequanal}
HD 51756 was observed by the CoRoT satellite during the second long run of the mission (LRa02) as part of the asteroseismology programme, while monitoring a field in the direction of the Galactic anticenter from 2454784.491822 HJD for almost 115 days (from 13  November 2008 to 8 March 2009). All flagged observations were removed from the light curve, leaving us with 278\,131 measurements (resulting in a duty cycle of $\approx90$\%). The decreasing trend of the CoRoT light curve has a well known instrumental origin \citep{2009A&A...506..411A}, hence it was also removed by dividing by a linear fit. As after this step there were no clear jumps or trends visible anymore, we used the resulting dataset in our analysis.

\begin{figure*}
\resizebox{\hsize}{!}{\includegraphics{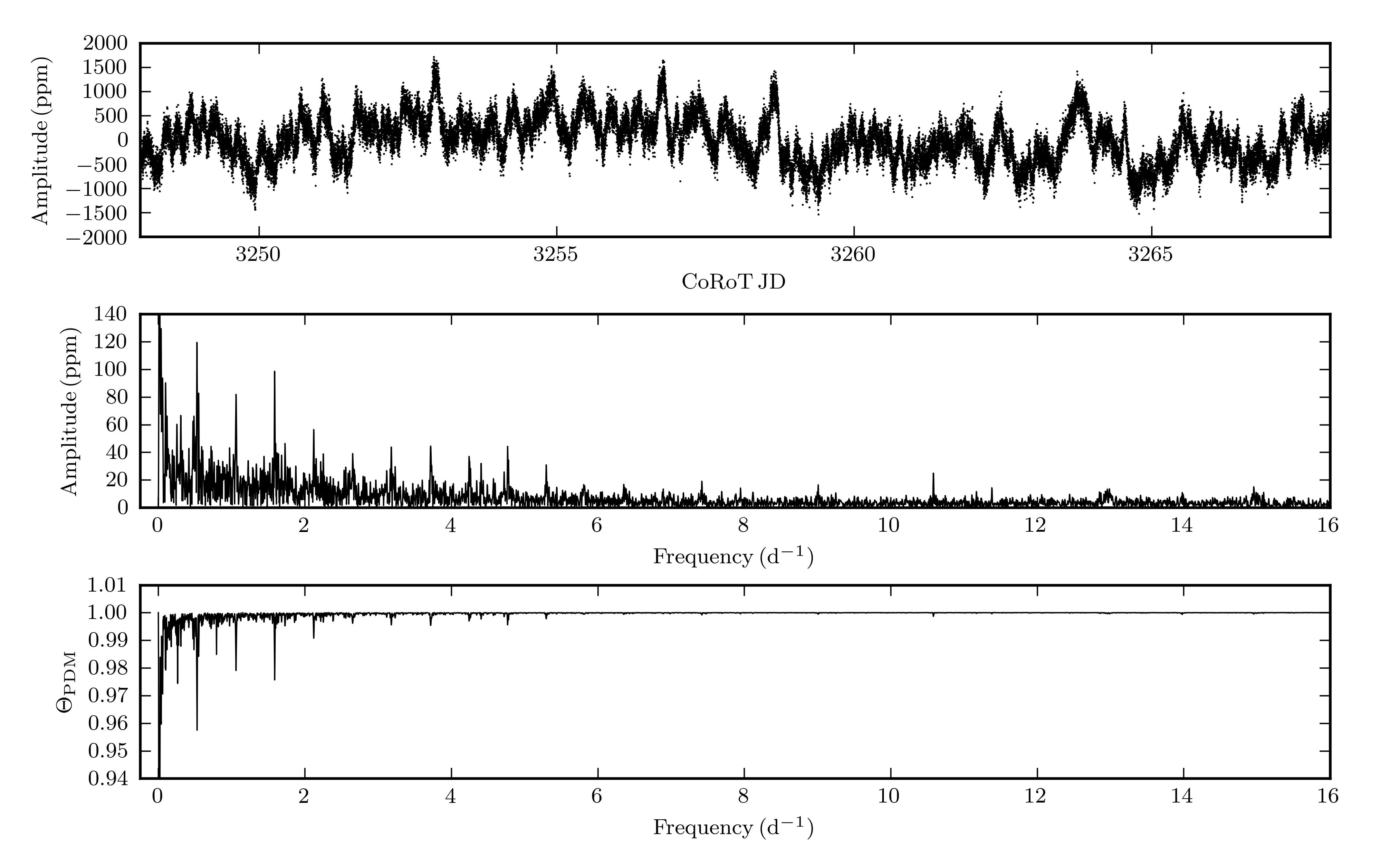}} 
\caption{(\textit{upper panel}) Part of the reduced CoRoT light curve, showing distinctive sharp recurring features every $\approx1.9$ days. This can be easily seen around 3255 CoRoT JD, where the occurrence of features causing the highest amplitude variations is clearly periodic. (\textit{middle panel}) The Scargle periodogram of the full CoRoT light curve showing a series of harmonics at integer multiple frequencies of the main peak at $f_0=0.52687\pm0.00004\mathrm{\,d}^{-1}$. (\textit{lower panel}) $\Theta_{PDM}$-statistics of the light curve, showing the same structure as the Scargle periodogram.}
\label{corotlcurvefourier}
\end{figure*}

The first visual inspection of the light curve showed a well traceable repetitive pattern over almost all the timespan of the observation, with a length of $\approx2$\,days (see Fig.\,\ref{corotlcurvefourier}). The shape and features of the light curve show a noteworthy resemblance to the O8.5\,V star HD\,46149 observed during the SRa02 short run of CoRoT and analysed in detail by \citet{2010A&A...519A..38D}.

We performed an iterative prewhitening procedure \citep[see e.g.][]{2009A&A...506..111D} using the traditional linear Scargle periodogram \citep{1982ApJ...263..835S} . This resulted in a list of amplitudes ($A_j$), frequencies ($f_j$), and phases ($\theta_j$), by which the light curve can be modelled via $n_j$ frequencies in the well-known form of \[F(t_i)=c+\sum_{j=1}^{n_f}A_j\sin[2\pi(f_j t_i + \theta_j)].\] We consider a peak significant if it exceeds an amplitude signal-to-noise ratio of 4 \citep[see][]{1993A&A...271..482B}. The noise level was calculated as the average amplitude -- before prewhitening -- in a $3\,\mathrm{d}^{-1}$ interval centered on the frequency of interest. Only 27 peaks of the Scargle periodogram (Fig.\,\ref{corotlcurvefourier}) met the signal-to-noise criterion, and using only the significant frequencies listed in Table\,\ref{frequtable}, we are unable to model the light curve on a satisfactory level. Even using all the 93 peaks which have an SNR above 3, the constructed model lacks the characteristics of the original dataset, leaving most of the non-sinusoidal sharp features in the residual light curve. The noise level in the original periodogram around 1.5--5--10$\,\mathrm{d}^{-1}$ is 20.7--7.8--3.7\,ppm, respectively, while it is 15.7--7.1--3.5 ppm in the periodogram of the residual light curve (which was created by subtracting a model with the parameters listed in Table\,\ref{frequtable}).

For comparison -- and because of the non-sinusoidal shape of the signal -- we also used the phase dispersion minimisation (PDM) method of \citet{1978ApJ...224..953S} to calculate the $\Theta_{PDM}$-statistics, which shows the same global structure and features as the Scargle periodogram (see Fig.\,\ref{corotlcurvefourier}).

\begin{figure}
\resizebox{\hsize}{!}{\includegraphics{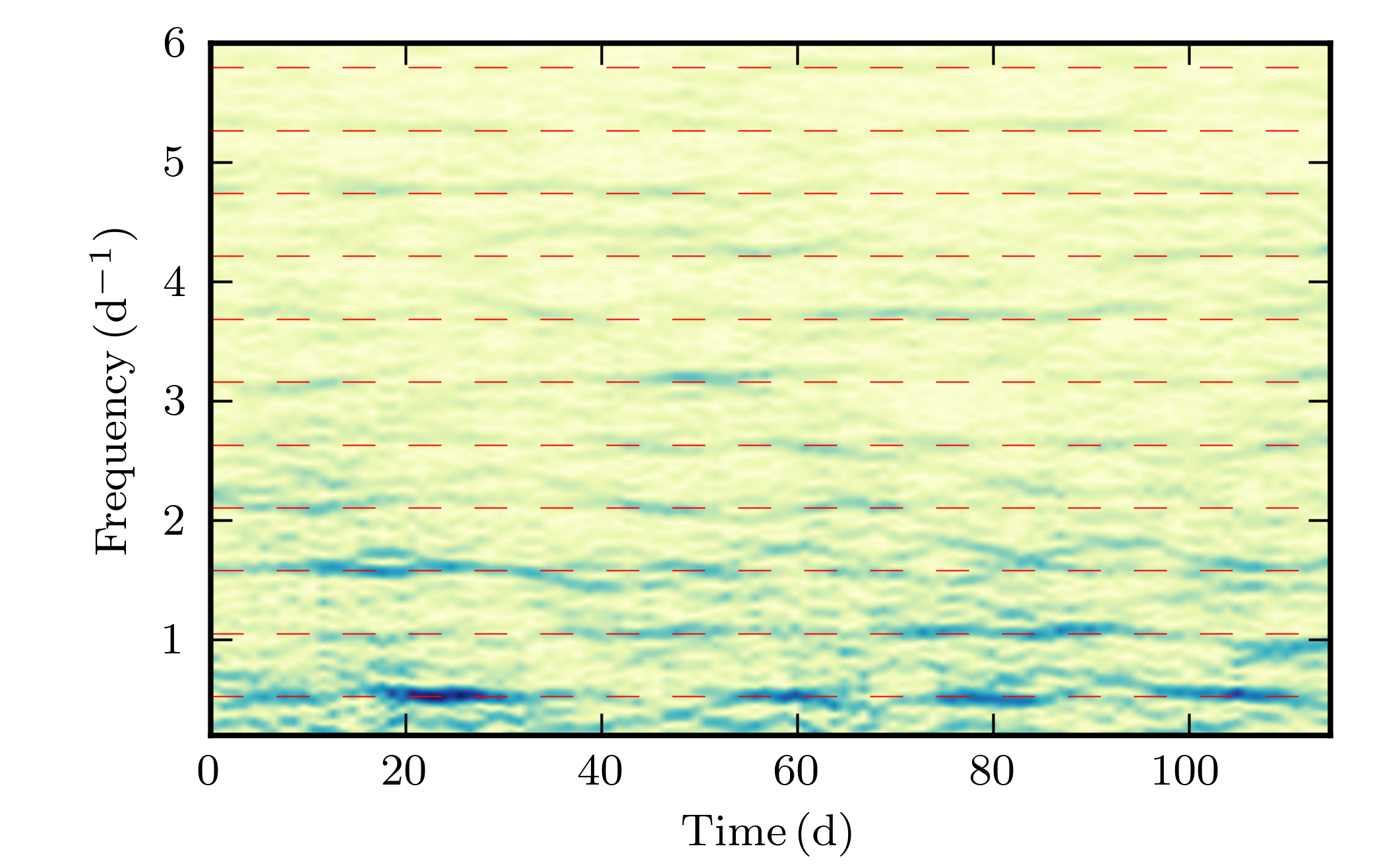}} 
\caption{Short-time Fourier transformation ($\mathrm{window}=5f_0$) of the $[0.2,\,6.0]\,\mathrm{d}^{-1}$ region of the light curve, showing that the amplitudes connected to different harmonics of the main peak at $f_0=0.52687\pm0.00004\mathrm{\,d}^{-1}$ (indicated as red dashed lines) vary significantly over time.}
\label{timefrequency}
\end{figure}

To check the stability of features in the frequency spectrum, time-resolved Fourier transformations were also calculated with different window widths. It is clearly visible from Fig.\,\ref{timefrequency} that the amplitudes connected to the significant frequencies vary significantly over the span of the time series.

Of the 27 frequencies with $\mathrm{SNR}\ge4$, 11 are connected with $f_0$ (Table\,\ref{frequtable}). All these frequency properties are a-typical for stellar oscillations in hot B stars and point towards another origin of the variability.

\subsection{Repetitive patterns}
The classical Fourier-techniques revealed a series of harmonics at integer multiple frequencies of the main peak at $f_0=0.52687\pm0.00004\mathrm{\,d}^{-1}$. All harmonics (using a notation of $f_0^n$ for the $n$th harmonic of $f_0$) can be traced in the power spectrum down to $f_0^{15}$ (with $\mathrm{SNR}\ge3$), plus $f_0^{17}$ and $f_0^{20}$ are also visible. There is a small but systematic difference between the observed $f_0^n$ and the exact $nf_0$ $(n\in\mathbb{Z})$ values. Fixing the frequency values of the harmonics to $nf_0$ and carrying out a nonlinear least-squares fitting procedure with these peaks gives a worse model, with remaining higher residual values caused by un-prewhitened peaks right next to the removed fixed frequency harmonics. The observed frequency values are always higher than the ones used in the fitting (fixed at integer multiples of $f_0$), and this difference increases with frequency (or $n$), starting with $\Delta f^2_0 = 2f_0-f_0^2 \approx -0.008\mathrm{\,d^{-1}}$ and continuously growing to $\Delta f^{20}_0 = 20f_0-f_0^{20} \approx -0.05\mathrm{\,d^{-1}}$. This might be a sign of rotational modulation with the rotational period of one of the components. If differential rotation were present, it could complicate the structure of the peaks and it might explain in part the differences between the observed peak values and exact multiples of $f_0$.

\begin{figure}
\resizebox{\hsize}{!}{\includegraphics{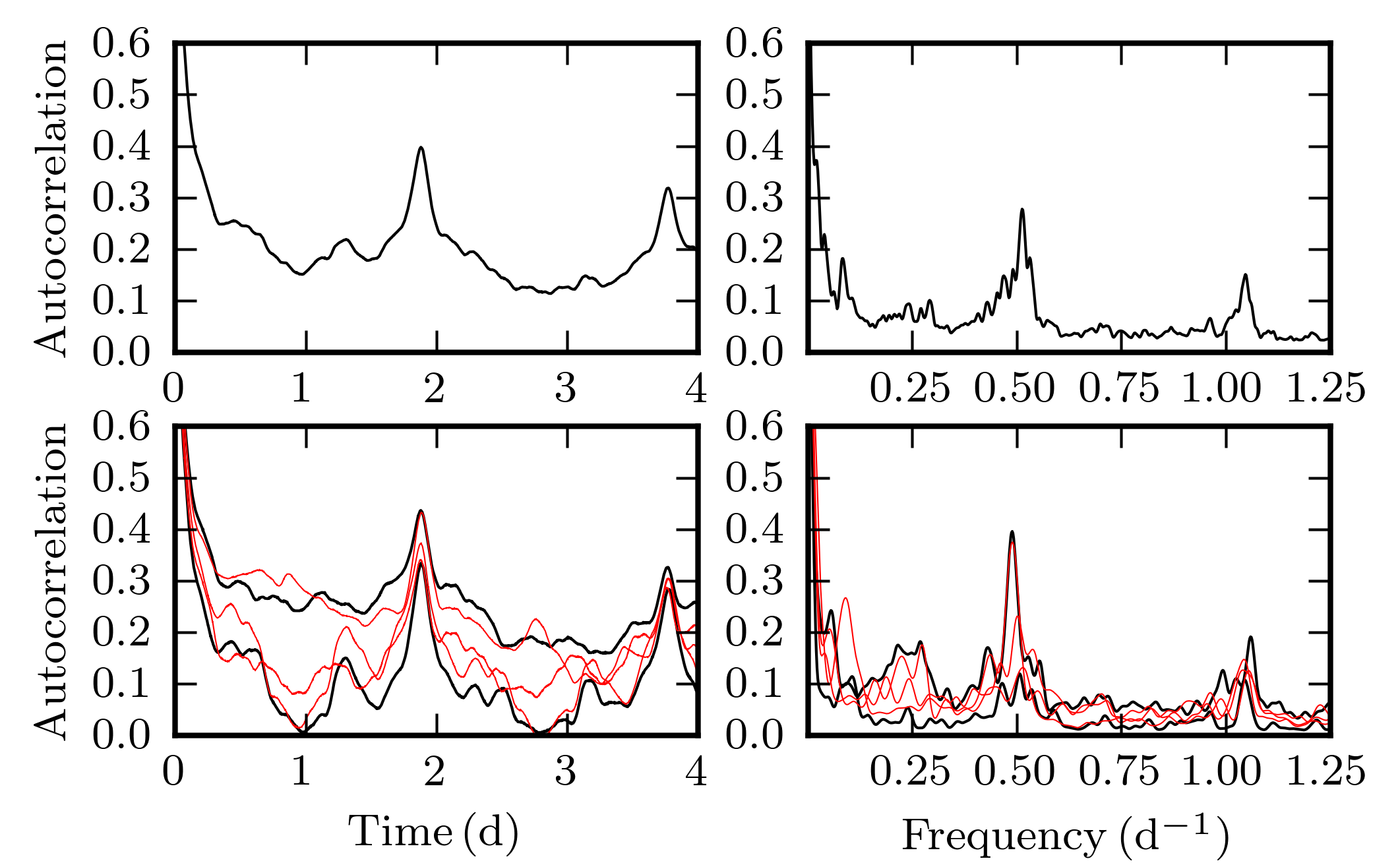}} 
\caption{Autocorrelation function of the CoRoT light curve (\textit{upper left}), of the power spectrum (\textit{upper right}), of the light curve cut into two and three pieces (\textit{lower left}), and of the power spectra which were calculated from the pieces (two and three, plotted with thicker and thinner lines, respectively) of the light curve  (\textit{lower right}).}
\label{autocorr}
\end{figure}

Another approach to search for spacings and recurrent features both in the light curve and the power spectrum is to use the autocorrelation function. We calculated the autocorrelation of the light curve, and the power spectrum (here the square of the Scargle periodogram). These tests were repeated after the dataset was cut into two, and three pieces, respectively. The results displayed on Fig.\,\ref{autocorr} show clear signs corresponding to the repetitive nature of the variations in the light curve. This signature stays quite clear even for shorter subsets of the data, but gets less significant in case of the power spectrum calculated from these data sets. Furthermore, the location of the first peak of the autocorrelation function stays almost perfectly the same even for three subsets, but this stability can not be seen in case of the autocorrelation of the power spectra of the subsets.

\subsection{The origin of repetitive variations}\label{origin}
A possible source of the variability might be atmospheric features, like spots or chemical inhomogeneities on the surface. In that case the observed brightness variations can be explained by the appearance and disappearance of such regions, either because of repeated creation or destruction or because of the rotation of the star. Here, we are in favour of the second option, because of the signal's self-similarity, and because -- as we show it below -- it is consistent with the determined $v \sin i$ value. For further details on how these features can actually produce the observed cusp-like peaks, we refer to \citet{2010_degroote_phd}.

\begin{figure*}
\resizebox{\hsize}{!}{\includegraphics{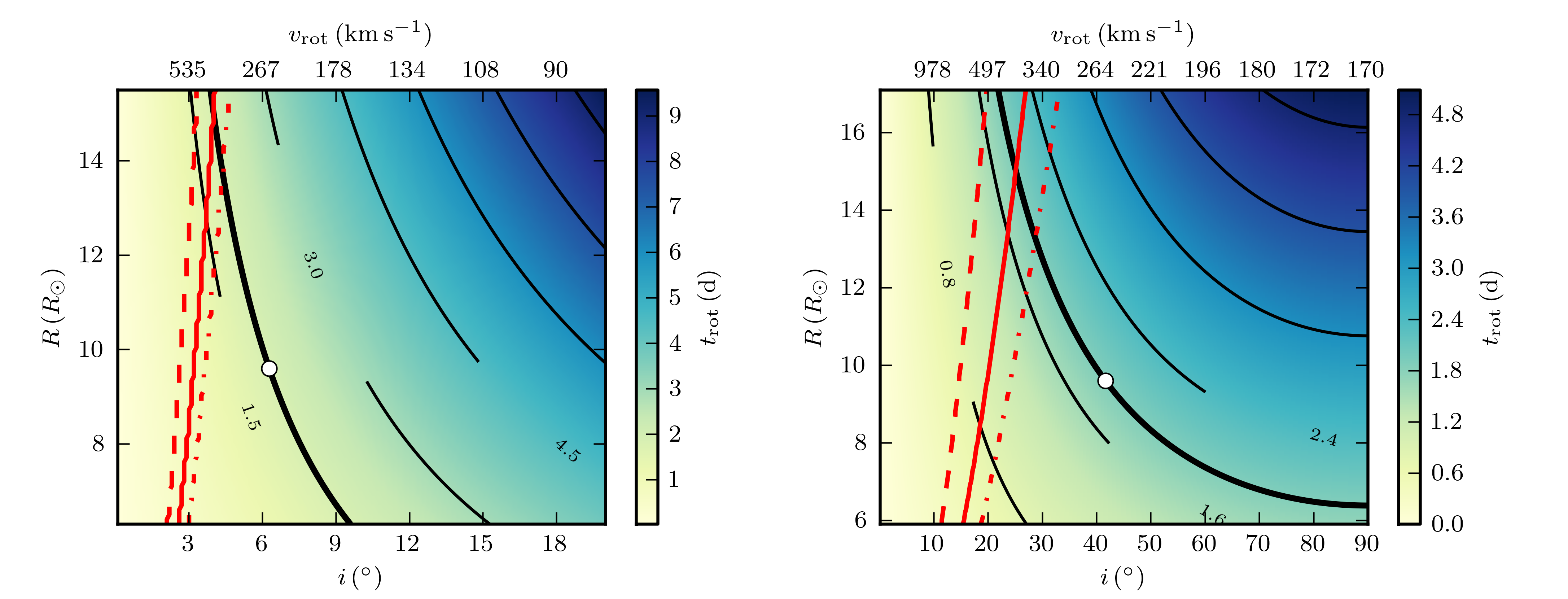}} 
\caption{Possible $i,R,v_\mathrm{rot}$ combinations in agreement with the observed $v\sin i=28\,\mathrm{km\,s}^{-1}$ of the primary (\textit{left}) and the $v\sin i=170\,\mathrm{km\,s}^{-1}$ of the secondary component (\textit{right}). Plots show (via colours and thin black curved contour lines) rotation periods calculated from the observed projected rotational velocities at given $i,R$ values. The thick contour line corresponds to the main peak at 1.898\,d from the frequency spectrum, and the dashed, continuous, and dot-dashed red lines show the break-up velocity (with speeds rising towards the left) for $\mathcal{M}_\mathrm{max}$, for $\mathcal{M}$, and for $\mathcal{M}_\mathrm{min}$, respectively. The thick contour line corresponds with the parameter-combinations which fit the hypothesis in Sect.\,\ref{origin}, and the white dot shows the location of the most probable $R$ (and thus $i$) value from Table\,\ref{gridsearchresults}.}
\label{rotationi}
\end{figure*}

Knowing the projected rotational velocities of both components, we test if it is physically possible that one of the components has a rotational period which is compatible with our hypothesis that the series of harmonics in the frequency spectrum is a sign of rotational modulation. To carry out this test, we calculated the inclination values which correspond to a rotational period of $1/f_0\,\mathrm{(1.898\,d)}$ within the error bars of the $R$ and $v \sin i$ values (from Table\,\ref{gridsearchresults}), and we checked if the equatorial rotational speed at this set-up was below the critical velocity for the given $R,\mathcal{M}$ combination. If we suppose that the primary is responsible for the repetitive features in the light curve, then its inclination has to be $i_\mathrm{A}=6.3^{+4.7}_{-3.0}$\,degrees (Fig.\,\ref{rotationi}). On the other hand, assuming that the variations are connected to the secondary, then we conclude an $i_\mathrm{B}=41.6^{+48.4}_{-21.7}$\,degrees and we end up with an equator-on solution already slightly before reaching the lowest possible $R$ value. Taking into account only the most probable $R$ values we end up with $i_\mathrm{A}=6.3^{+0.9}_{-0.9}$\,degrees and $i_\mathrm{B}=41.6^{+4.7}_{-4.4}$\,degrees.

As it is clearly visible on Fig.\,\ref{rotationi}, these setups may result in exceeding the break-up velocity if the radius of the given component is in the very upper part of the error box. Lowering the mass, the radius limit also drops. We can conclude that most of the possible configurations are indeed compatible with our hypothesis of a rotational origin of the variability.

\subsection{Inferences from the non-detection of pulsation}
For all the models passing through the error boxes of the primary and secondary indicated in Fig.\,\ref{grids1}, we considered the frequencies of the fundamental radial mode and of the few lowest radial overtones, as well as the lowest-order p and g axisymmetric modes of degree 1 and 2. None of those eigenfrequencies come close to $f_0$. It is thus very unlikely that $f_0$, with so many harmonics present in the light curve, results from a pulsation mode, as one does not expect a high-overtone mode or a high-degree mode to show so many harmonics of its frequency \citep[e.g.,][]{2009A&A...506..111D}.

\begin{figure*}
\resizebox{\hsize}{!}{\includegraphics{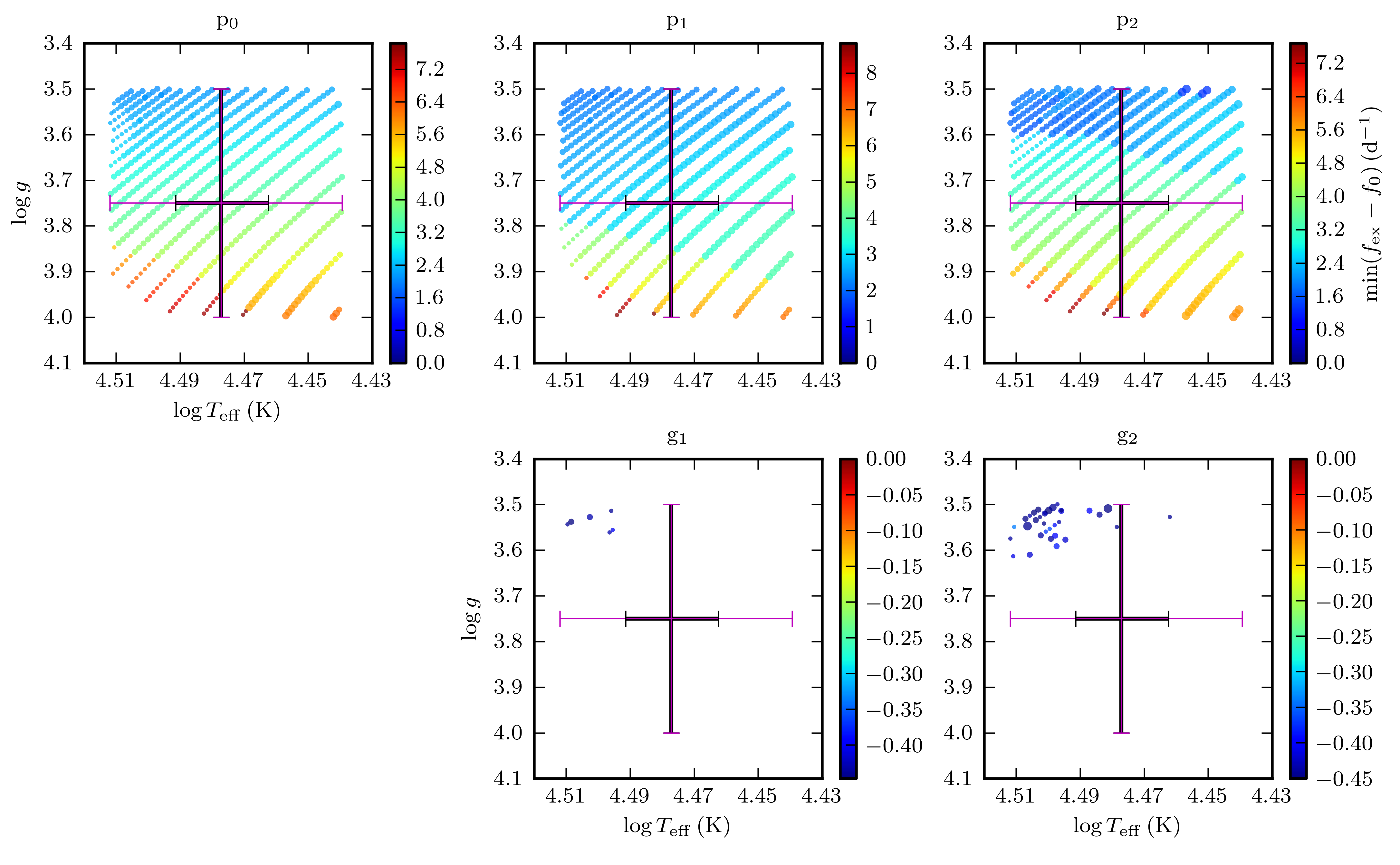}} 
\caption{$\log T_\mathrm{eff}$--$\log g$ diagrams showing an overview of models where different modes are predicted to be excited in our non-adiabatic computations for the error box of the components of HD\,51756. Each filled circle represents a model along the evolutionary tracks where modes are predicted to be excited. The size of the symbols refers to the number of excited modes ($1-4$ per mode type), while the colour corresponds to the frequency difference between $f_0$ and the closest mode in the model. The spectroscopic parameters and corresponding error estimations of HD\,51756 are indicated by thick and thin crosses for the primary and secondary component, respectively.}
\label{mad}
\end{figure*}

We made non-adiabatic computations \citep[with the code MAD developed by][]{2002A&A...385..563D} for the mentioned models to check which modes are predicted to be excited by the present input physics and theory of mode excitation (see Fig.\,\ref{mad}). We need to understand why the star does not oscillate, as the theoretical calculations predict p modes to be excited throughout the error box of HD\,51756, except the very bottom left region. Typical frequencies are $3.5-6.5\,\mathrm{d}^{-1}$ around the center of our error box. This range extends until $1.9\,\mathrm{d}^{-1}$ and $9.4\,\mathrm{d}^{-1}$ towards the lower and higher gravity values, respectively. Moreover, g modes are also expected to be excited in some of the higher mass models, with frequencies near $0.1\,\mathrm{d}^{-1}$.

Finally, we should remember that CoRoT measured the integrated flux from two stars, therefore the observed amplitude of the rotational modulation is damped, and this would be true also for the amplitudes of the pulsational modes, if any.

\section{Conclusions}
We analysed high-resolution spectroscopy and high-quality space based photometry of HD\,51756. The former -- via synthetic spectrum fitting and matching the results with evolutionary tracks -- enabled us to derive the fundamental stellar parameters of both components of the underlying double lined spectroscopic binary, while the latter led us to conclude that none of these stars show pulsations. We interpret the repetitive patterns of the light curve, the observed harmonic structure in the frequency spectrum, and the strong, stable peaks in the autocorrelation functions as signs for rotational modulation. This hypothesis is compatible with the observed and deduced stellar parameters.

The absence of oscillations in the components of HD\,51756 requires an explanation. It was recently found by Balona et al.\,(in preparation) that some stars observed with the Kepler space mission, which are also monitored at $\mu$mag precision and situated in the $\beta$\,Cep instability strip, do not pulsate either. A mixture of non-pulsators and pulsators with or without rotational modulation was already established from ground-based data for stars in the g-mode instability strip of the cooler slowly pulsating B stars \citep[e.g.,][]{2004A&A...413..273B}. It was interpreted as due to a difference in magnetic field strength and age, the Bp stars being younger and having stronger fields than the pulsators, although this interpretation is still a matter of debate \citep[e.g.,][]{2007A&A...466..269B, 2009MNRAS.398.1505S}. As far as we are aware, no magnetic field measurements are available for HD\,51756.


\begin{acknowledgements}
The research leading to these results has received funding from the European Research Council under the European Community's Seventh Framework Programme (FP7/2007--2013)/ERC grant agreement n$^\circ$227224 (PROSPERITY), as well as from the Research Council of K.U.Leuven grant agreement GOA/2008/04. EN acknowledges financial support of the NN203 302635 grant from the MNiSW. EP and MR acknowledge financial support from the Italian ESS project, contract ASI/INAF I/015/07/0, WP 03170. This research has made use of the Washington Double Star Catalog maintained at the U.S. Naval Observatory.
\end{acknowledgements}

\bibliographystyle{aa}
\bibliography{HD51756}


\Online
\begin{appendix}
\section{Tables}\label{tables}

Table\,\ref{positions} contains all the position measurements of HD\,51756 (catalogued as BU\,327, but widely known as WDS\,J06585-0301) found in the Washington Double Star Catalog. Error estimates -- when available -- can be found in the original references.


Table\,\ref{frequtable} lists the Fourier parameters of significant peaks in the periodogram.

\begin{table}
\caption{Position measurements of HD 51756 AB}
\label{positions}
\centering
\renewcommand{\arraystretch}{0.75}
\begin{tabular}{l c l l c c}
\hline\hline
Date & $\theta(\degr)$ & $\rho(\arcsec)$ & N & Method\tablefootmark{a} &  Reference\tablefootmark{b} \\
\hline
1875.06  &  90    &  1    & 1 &    A   &  1 \\  
1876.83  & 100.8  &  0.96 & 2 &    A   &  D1883 \\  
1878.12  &  97.2  &  n/a  & 1 &    A   &  2 \\  
1881.05  &  96.1  &  0.75 & 1 &    A   &  Bu1883 \\  
1888.10  &  95.8  &  0.73 & 2 &    A   &  3 \\  
1888.222 &  95.5  &  0.84 & 4 &    B   &  4 \\  
1892.08  &  95.8  &  0.79 & 3 &    A   &  Bu1894 \\  
1894.40  &  98.1  &  0.64 & 5 &    A   &  Sp1909 \\  
1899.01  &  93.6  &  0.59 & 1 &    A   &  Bu1900 \\  
1902.544 &  93.9  &  0.68 & 6 &    A   &  5 \\ 
1903.14  & 100.0  &  0.64 & 2 &    A   &  9 \\ 
1906.16  &  95.9  &  0.87 & 3 &    A   &  6 \\ 
1906.17  & 101.5  &  0.81 & 1 &    A   &  Frm1907 \\ 
1909.06  & 100.4  &  0.90 & 1 &    A   &  7 \\ 
1909.06  & 101.3  &  0.88 & 1 &    A   &  7 \\ 
1909.13  &  96.9  &  0.70 & 2 &    A   &  Wz1912 \\ 
1911.15  &  91.7  &  0.56 & 2 &    A   &  8 \\ 
1920.09  &  98.1  &  0.77 & 1 &    A   &  10 \\ 
1924.35  &  99.4  &  0.83 & 3 &    A   &  10 \\ 
1924.79  &  97.5  &  0.94 & 1 &    A   &  10 \\ 
1930.57  &  98.3  &  0.71 & 2 &    A   &  Bon1938 \\ 
1930.84  &  96.2  &  0.80 & 3 &    A   &  11\\
1935.43  & 102.0  &  0.78 & 4 &    A   &  12\\ 
1942.88  &  95.1  &  0.62 & 3 &    A   &  13 \\ 
1945.31  & 103.6  &  0.68 & 2 &    B   &  VBs1954 \\ 
1947.17  & 102.8  &  0.76 & 2 &    A   &  VBs1954 \\ 
1953.22  &  98.5  &  0.81 & 6 &    A   &  Rab1961b\\ 
1956.11  &  98.2  &  0.74 & 2 &    A   &  14 \\ 
1961.04  & 101.6  &  0.67 & 4 &    A   &  15\\ 
1965.034 & 100.1  &  0.69 & 4 &    A   &  16 \\ 
1966.182 & 101.4  &  0.58 & 3 &    A   &  17 \\ 
1968.045 & 101.4  &  0.66 & 3 &    A   &  18\\ 
1974.717 & 101.1  &  0.59 & 3 &    A   &  19\tablefootmark{c}\\ 
1975.522 & 101.6  &  0.55 & 3 &    B   &  Wak1985 \\ 
1991.25  & 102.   &  0.71 & 1 &    T   &  20\\ 
1991.92  & 104.0  &  0.68 & 1 &    T   &  21 \\ 
1994.129 & 101.2  &  0.67 & 1 &    S   &  22\\ 
1994.194 &  99.7  &  0.63 & 1 &    S   &  22\\ 
1994.194 & 100.5  &  0.68 & 1 &    S   &  22\\ 
1995.188 & 100.6  &  0.66 & 1 &    S   &  22\\ 
1995.199 & 103.0  &  0.68 & 1 &    S   &  22\\ 
1995.202 & 102.1  &  0.70 & 1 &    S   &  22\\ 
1995.202 & 101.8  &  0.67 & 1 &    S   &  22\\ 
1995.205 & 101.5  &  0.69 & 1 &    S   &  22\\ 
1995.207 & 101.9  &  0.63 & 1 &    S   &  22\\ 
1996.8960& 101.5  &  0.695& 1 &    S   &  23\\ 
1997.184 & 101.5  &  0.67 & 1 &    S   &  24\\ 
1997.184 & 102.2  &  0.69 & 1 &    S   &  24\\ 
\hline
\end{tabular}
\tablefoot{\tablefoottext{a}{Methods of observations. A: Refractor, micrometer; B: Reflector, micrometer; T: Hipparcos or Tycho observation; S: Speckle interferometry.}\tablefoottext{b}{For special references - which are not in the SAO/NASA Astrophysics Data System - see the References and discoverer codes of The Washington Double Star Catalog at http://ad.usno.navy.mil/wds/Webtextfiles/wdsnewref.txt}\tablefoottext{c}{These observations were made after the closing of the referred article, but before the retirement of Alan Behall. Observing techniques, instrumentation and particulars are the same as described by \citet{1976PUSNO..24....1B}}}
\tablebib{(1) \citet{1875AN.....86..337B}; (2) \citet{1879PCinO...5....1S}; (3) \citet{1930QB821.L4.......}; (4) \citet{1890AN....125..225T}; (5) \citet{1905PFAO....2c...1D}; (6) \citet{1907AN....174..209O}; (7) \citet{1909AN....182..253O}; (8) \citet{1912MNRAS..73...93E}; (9) \citet{1904MNRAS..64..789.}; (10) \citet{1928AN....233..393O}; (11) \citet{1932PFAO....5a...1O}; (12) \citet{1938JO.....21..161B}; (13) \citet{1955JO.....38..109V}; (14) \citet{1957AJ.....62..153W}; (15) \citet{1963AJ.....68..114W}; (16) \citet{1966PUSNO..18....1W}; (17) \citet{1969PUSNO..22....1W}; (18) \citet{1972PUSNO..22....2W}; (19) \citet{1976PUSNO..24....1B}; (20) \citet{1997yCat.1239....0E}; (21) \citet{2002A&A...384..180F}; (22) \citet{1999AJ....117.1905G}; (23) \citet{2001AJ....121.1597H}; (24) \citet{1999AJ....118.1395D}.}
\end{table}

\begin{table*}
\caption{Table of Fourier parameters (frequencies ($f_j$), amplitudes ($A_j$), and phases ($\theta_j$)) of peaks with a signal-to-noise ratio (SNR) $\ge4$, plus clear harmonics ($f_0^n$) of the main peak at $f_0=0.52687\pm0.00004\mathrm{\,d}^{-1}$ down to $\mathrm{SNR}\ge3$. SNR is calculated over $3\,\mathrm{d}^{-1}$ in the periodogram.}
\label{frequtable}
\centering
\begin{tabular}{c c c c c c c c}
\hline\hline
$f\,(\mathrm{d}^{-1})$ & $\epsilon_f\,(\mathrm{d}^{-1})$ & $A\,(\mathrm{ppm})$ & $\epsilon_A\,(\mathrm{ppm})$ & $\theta\,(2\pi/\mathrm{rad})$ & $\epsilon_{\theta}\,(2\pi/\mathrm{rad})$ & SNR & note\\
\hline
  0.009071 &   0.000035 &   167.4 &     1.1 &   0.2563 &   0.0073 &  10.4 &  \\
  0.017911 &   0.000030 &   183.8 &     1.1 &  -0.1304 &   0.0063 &  12.2 &  \\
  0.033302 &   0.000042 &   137.5 &     1.1 &  -0.0228 &   0.0087 &   8.5 &  \\
  0.040540 &   0.000060 &    80.1 &     1.0 &   0.3908 &   0.0124 &   6.0 &  \\
  0.062041 &   0.000057 &    89.8 &     1.0 &   0.2046 &   0.0118 &   6.4 &  \\
  0.098717 &   0.000058 &   102.2 &     1.0 &  -0.0143 &   0.0121 &   6.2 &  \\
  0.119596 &   0.000080 &    58.2 &     0.9 &  -0.0415 &   0.0165 &   4.5 &  \\
  0.137555 &   0.000083 &    59.1 &     0.9 &   0.4332 &   0.0173 &   4.3 &  \\
  0.253269 &   0.000075 &    60.2 &     0.9 &   0.3487 &   0.0156 &   4.7 &  \\
  0.306826 &   0.000071 &    63.9 &     1.0 &  -0.3190 &   0.0146 &   5.1 &  \\
  0.472920 &   0.000071 &    65.3 &     1.0 &  -0.2757 &   0.0147 &   5.0 &  \\
  0.485587 &   0.000081 &    69.5 &     0.9 &  -0.2933 &   0.0168 &   4.4 &  \\
  0.496949 &   0.000088 &    47.8 &     0.9 &   0.1944 &   0.0182 &   4.1 &  \\
  0.526865 &   0.000042 &   124.1 &     1.0 &  -0.3897 &   0.0088 &   8.5 &  $f_0$\\
  0.537154 &   0.000075 &    61.5 &     0.9 &  -0.0783 &   0.0155 &   4.8 &  \\
  0.551246 &   0.000060 &    69.7 &     1.0 &   0.0598 &   0.0124 &   6.0 &  \\
  1.061396 &   0.000058 &    94.0 &     1.0 &  -0.3826 &   0.0120 &   6.2 &  $f_0^2$\\
  1.588351 &   0.000050 &   101.1 &     1.0 &   0.3509 &   0.0104 &   7.2 &  $f_0^3$\\
  2.122824 &   0.000081 &    58.3 &     0.9 &  -0.3705 &   0.0167 &   4.4 &  $f_0^4$\\
  2.653912 &   0.000106 &    39.1 &     0.9 &  -0.0592 &   0.0219 &   3.4 &  $f_0^5$\\
  3.181300 &   0.000098 &    42.5 &     0.9 &   0.2905 &   0.0203 &   3.8 &  $f_0^6$\\
  3.718828 &   0.000094 &    45.3 &     0.9 &  -0.1821 &   0.0195 &   4.3 &  $f_0^7$\\
  4.243161 &   0.000110 &    37.2 &     0.8 &  -0.1519 &   0.0229 &   4.1 &  $f_0^8$\\
  4.769149 &   0.000094 &    45.5 &     0.9 &   0.1209 &   0.0195 &   5.5 &  $f_0^9$\\
  5.297008 &   0.000119 &    31.9 &     0.8 &  -0.1407 &   0.0248 &   4.6 &  $f_0^{10}$\\
  5.809200 &   0.000188 &    17.3 &     0.6 &   0.0690 &   0.0390 &   3.1 &  $f_0^{11}$\\
  6.354499 &   0.000185 &    16.4 &     0.6 &  -0.3715 &   0.0384 &   3.2 &  $f_0^{12}$\\
  6.894291 &   0.000213 &    13.5 &     0.6 &   0.4758 &   0.0442 &   3.1 &  $f_0^{13}$\\
  7.421695 &   0.000163 &    19.8 &     0.7 &  -0.2722 &   0.0338 &   4.2 &  $f_0^{14}$\\
  7.949167 &   0.000198 &    14.9 &     0.6 &  -0.1871 &   0.0411 &   3.6 &  $f_0^{15}$\\
  9.009843 &   0.000178 &    17.3 &     0.6 &  -0.4902 &   0.0369 &   4.6 &  $f_0^{17}$\\
 10.583197 &   0.000140 &    24.9 &     0.7 &  -0.3271 &   0.0291 &   7.4 &  $f_0^{20}$\\
 13.969336 &   0.000292 &     9.0 &     0.5 &   0.1131 &   0.0606 &   4.0 &  \tablefootmark{a}\\
\hline
\end{tabular}
\tablefoot{The group of low frequency values show up because of noise and not completely removed trends. Several frequencies close to $f_0$ can be explained by the fact that neither the frequency nor the amplitude of $f_0$ is stable, as it can be clearly seen on Fig.\,\ref{timefrequency}. \tablefoottext{a}{Peak corresponding to the orbital period of the CoRoT satellite.}}
\end{table*}

\end{appendix}

\end{document}